\documentstyle[epsf,sprocl]{article}
\epsfverbosetrue

\def\Journal#1#2#3#4{{#1} {\bf #2}, #3 (#4)}

\def\NPB{{\em Nucl. Phys.} B}
\def\PLB{{\em Phys. Lett.}  B}
\def\PRL{\em Phys. Rev. Lett.}
\def\PRD{{\em Phys. Rev.} D}
\def\ZPC{{\em Z. Phys.} C}
\def\PTP{\em Prog. Theor. Phys.}

\def\be{\begin{equation}}
\def\ee{\end{equation}}
\def\bea{\begin{eqnarray}}
\def\eea{\end{eqnarray}}

\begin{document}

\title{TOP QUARK PHYSICS \\ FOR BEAUTIFUL AND CHARMING PHYSICISTS} 

\author{S.~WILLENBROCK}

\address{Department of Physics, University of Illinois, 1110 West Green 
Street, \\ Urbana, IL 61801}


\maketitle\abstracts{I discuss two aspects of top-quark physics:
$V_{tb}$ and the top-quark mass.  The similarities and differences with
bottom and charm physics are emphasized.}

While preparing this talk, I was struck by how different the physics of 
the top quark is from that of bottom and charm.  This is largely a 
consequence of the very short lifetime of the top quark, $\Gamma^{-1} 
\approx (1.5 \;\rm GeV)^{-1}$, less than the characteristic time scale of 
nonperturbative QCD, $\Lambda_{\rm QCD}^{-1} \approx (200 \;\rm MeV)^{-1}$.  
As a result, the physics of the top quark is free of the effects of 
nonperturbative QCD.  Thus one is not concerned with form factors, 
decay constants, exclusive decays, 
and other such topics which one usually associates with heavy-flavor physics.
The lack of nonperturbative effects allows for precision studies of the 
top quark, which gives us access to interesting physics both within 
the standard model and, hopefully, beyond.

Within the standard model, there are only a few parameters associated 
with the top quark: $m_t, V_{tb}, V_{ts}$, and $V_{td}$.\footnote{Assuming
three generations, $V_{ts}$ and $V_{td}$ are determined indirectly from
loop processes in $B$ and $K$ physics.\cite{BF}}  One goal of 
top-quark physics is to measure these parameters precisely.  However, we 
have much higher hopes for top-quark physics; we hope that it will reveal 
physics beyond the standard model.  

In this talk I will consider top physics within the standard model.  If 
we are to find physics beyond the standard model, it is 
essential that we first understand top physics within the standard model.  
Furthermore, I have chosen to concentrate on two topics which I believe to
be of particular interest to heavy-flavor physicists.  
These topics are $V_{tb}$, and the measurement of $m_t$ from the $t\bar t$
threshold in lepton colliders.  
   
\section{$V_{tb}$}

It is amusing that, within the context of three generations, $V_{tb}$ is 
the best-known CKM matrix element (as a percentage of its value), 
$V_{tb} = .9989 - .9993$,\cite{PDG} 
despite the fact that it has never been directly measured.  We owe this 
to beauty physicists, who have measured $V_{cb}$ and $V_{ub}$ to be small:
three-generation unitarity then implies that $V_{tb}$ must be near unity.
The direct measurement of $V_{tb}$ is therefore only of interest if we 
entertain the possibility of more than three 
generations.\footnote{We know from LEP that if there is a fourth generation, 
the associated neutrino is not light; however, this does not rule out a 
fourth generation.}
In this case 
$V_{tb}$ is almost entirely unconstrained, $V_{tb} = 0 - .9993$.\cite{PDG}

We heard in the previous talk \cite{N} that CDF has measured 
$V_{tb} > .76$ ($95\%$ C.~L.), assuming three generations.  It is 
important to realize that this bound disappears if there are more than 
three generations.  Recall that CDF measures the ratio of the branching 
ratio of the top quark to bottom quarks over the branching ratio to all 
quarks,
\begin{equation}
\frac{BR(t\to bW)}{BR(t\to qW)} 
= \frac{|V_{tb}|^2}{|V_{td}|^2 + |V_{ts}|^2 +|V_{tb}|^2} = .99\pm .29 \;.
\end{equation}
If there are just three generations, then the denominator in the second 
expression is unity, and one obtains the lower bound on $V_{tb}$ 
quoted above.  However, if there are more than three generations, there 
is no bound on $V_{tb}$ from this measurement; one only learns that 
$V_{tb} >> V_{ts},V_{td}$.\footnote{Furthermore, if there are more than 
three generations, there is no lower bound on $V_{ts}$ or 
$V_{td}$.\cite{PDG}}  This point is understood instantly by beauty 
physicists; the fact that the bottom quark decays predominantly to charm 
does not determine $V_{cb}$, only that $V_{cb} >> V_{ub}$.

\subsection{Indirect information on $V_{tb}$}

Before we proceed to the direct measurement of $V_{tb}$, let's explore what we 
know about it indirectly.  Consider first precision electroweak 
measurements, such as the $\rho$ parameter,\cite{V}
\begin{equation}
\rho \equiv \frac{M_W^2}{M_Z^2 \cos^2\theta_W} = 1 + \frac{3G_F}{8\sqrt 2 
\pi^2}m_t^2
\label{rho}
\end{equation}
where the last term comes from the loop diagram in Fig.~1(a).  
The measurement of this parameter told us that the top quark mass must be 
less than about 200 GeV long before it was discovered.  If there were a 
fourth generation, then it would also contribute to the $\rho$ parameter, via 
the loop diagram in Fig.~1(b).  Recall that the loop correction to the 
$\rho$ parameter depends on the mass 
splitting within an SU(2) doublet.  Since the $tb$ doublet already 
contributes as much as is necessary to yield a $\rho$ parameter 
consistent with experiment, the mass splitting in the fourth generation 
must be small, $m_{t^\prime} \approx m_{b^\prime}$. Allowing for mixing 
between the third and fourth generations, Eq.~(\ref{rho}) becomes
\begin{equation}
\rho \approx 1 + \frac{3G_F}{8\sqrt 2 \pi^2}
\left[m_t^2 |V_{tb}|^2 + m_{t^\prime}^2 |V_{t^\prime b}|^2 \right]\;.
\label{rho2}
\end{equation}
The consequences of Eq.~(\ref{rho2}) are laid bare if we invoke 
four-generation unitarity to write $|V_{tb}|^2 = 1 - \epsilon^2, 
|V_{t^\prime b}|^2 = \epsilon^2$.\footnote{I am assuming $V_{ts}, V_{td} 
<< 1$ throughout this argument, for simplicity. Relaxing this assumption does
not change the conclusion of the argument.}  We then obtain
\begin{equation}
\rho \approx 1 + \frac{3G_F}{8\sqrt 2 \pi^2}\left[m_t^2 
     +\epsilon^2(m_{t^\prime}^2-m_t^2)\right]
\label{rho3}
\end{equation}
What we 
learn from Eq.~(\ref{rho3}) is that either $\epsilon^2$ is small, which 
corresonds to $V_{tb} \approx 1$, or $m_{t^\prime} \approx m_t$.  The 
lesson is that in order for $V_{tb}$ to deviate significantly from unity, 
there must be a fourth generation of quarks whose mass is not far above 
that of the top quark.

\begin{figure}
\begin{center}
\epsfxsize= 3.5in   
\leavevmode         
\epsfbox{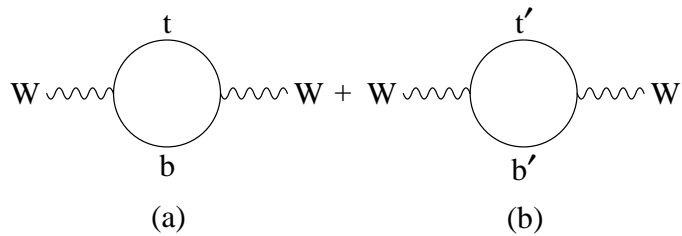}
\end{center}
\caption{One-loop contribution to the $\rho$ parameter from the (a) third 
and (b) fourth generation of quarks.}
\end{figure}

As beauty physicists know well, the top quark also contributes indirectly to 
flavor-changing neutral current processes such as $b \to s\gamma$, as 
shown in Fig.~2(a).  The observed rate for $b\to s 
\gamma$ is consistent with $V_{tb}\approx 1$.
If there were a fourth generation, it would also contribute to
$b\to s \gamma$, via Fig.~2(b).\cite{HSS}  
The total contribution to the amplitude from 
the third and fourth generation is proportional to 
\begin{equation}
A \sim F_2(m_t^2/M_W^2) V_{tb}V_{ts}^* 
+ F_2(m_t^{\prime 2}/M_W^2) V_{t^\prime b}V_{t^\prime s}^* 
\label{bsg}
\end{equation} 
where $F_2(x)$ is the Inami-Lim function.\cite{IL}  
Using the four-generation unitarity relation 
$V_{t^\prime b}V_{t^\prime s}^* + V_{tb}V_{ts}^* + V_{cb}V_{cs}^* + 
V_{ub}V_{us}^* = 0$, and neglecting the last term in this relation
(known to be small), we can rewrite Eq.~(\ref{bsg}) as 
\begin{equation}
A \sim - F_2(m_t^2/M_W^2) V_{cb}V_{cs}^* +
[F_2(m_t^{\prime 2}/M_W^2) - F_2(m_t^2/M_W^2)] V_{t^\prime b}V_{t^\prime s}^* 
\;.
\label{bsg2}
\end{equation} 
The first term above is all that is needed to produce a rate for $b\to s\gamma$
consistent with experiment.  The second term is negligible if either
$m_t^\prime \approx m_t$, or $V_{t^\prime b}V_{t^\prime s}^*$ is small, 
regardless of the value of $V_{tb}$.  Thus this process cannot constrain 
$V_{tb}$ if there are more than three generations.

\begin{figure}
\begin{center}
\epsfxsize= 3.5in
\leavevmode
\epsfbox{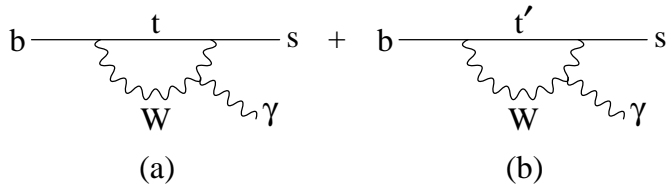}
\end{center}
\caption{One-loop contribution to $b\to s \gamma$ from the (a) third 
and (b) fourth generation of quarks.}
\end{figure}

We conclude that, although $V_{tb}\approx 1$ is consistent with all known 
phenomenology, the case is not airtight.  We should therefore measure 
it directly, and keep our eyes and minds open.

\subsection{Direct measurement of $V_{tb}$}

The best way to measure $V_{tb}$ at hadron colliders is via 
single-top-quark production.  There are two separate processes: 
quark-antiquark annihilation,\cite{CP} shown in Fig.~3(a), and 
$W$-gluon fusion,\cite{DW} shown in Fig.~3(b).  Both proceed via the weak 
interaction, with a cross section proportional to $|V_{tb}|^2$.  

A strategy to extract $V_{tb}$ from single-top-quark production is as 
follows.  One measures the cross section times the branching ratio of 
$t\to bW$, 
\begin{equation}
\sigma=\sigma(t\bar b) BR(t\to bW) \;.
\label{sigtb}
\end{equation}
To determine $BR(t\to bW)$, one uses the measured cross section for $t\bar 
t$ production and decay,
\begin{equation}
\sigma=\sigma(t\bar t) [BR(t\to bW)]^2\;.
\end{equation}
One extracts $BR(t\to bW)$ from this measurement, and inserts it into 
Eq.~(\ref{sigtb}) to obtain $\sigma(t\bar b)$, which is proportional to 
$|V_{tb}|^2$.  This procedure requires an accurate theoretical calculation 
of both  $\sigma(t\bar b)$ and $\sigma(t\bar t)$.  The former has been 
calculated to next-to-leading-order in the strong interaction for both
quark-antiquark annihilation \cite{SmW} and $W$-gluon fusion.\cite{SSW}
The latter has also been 
calculated to next-to-leading-order in the strong interaction, and is known
to $\pm 10\%$.\cite{NDE,CMNT}  The next-to-next-to-leading-order 
correction for all three processes would be desirable.\footnote{There 
are attempts to go beyond next-to-leading order by summing the effects of 
soft gluons.\cite{CMNT,LSV}}

\begin{figure}
\begin{center}
\epsfxsize= 3in
\leavevmode
\epsfbox{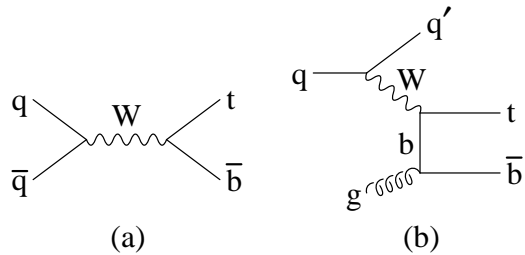}
\end{center}
\caption{Single-top-quark production via (a) quark-antiquark annihilation 
and (b) $W$-gluon fusion.}
\end{figure}

The quark-antiquark process has the advantage that the quark distribution 
functions are well known, while the $W$-gluon fusion process suffers from 
the uncertainty in the gluon distribution function.  Furthermore, one can 
exploit the similarity of the quark-antiquark-annihilation process to 
Drell-Yan to measure the incident quark-antiquark flux.  However, $W$-gluon 
fusion has the advantage of a larger cross section at the Tevatron, and a 
much larger cross section at the LHC; it is doubtful that the 
quark-antiquark-annihilation process will be observable above
backgrounds at the LHC.  Thus the two processes are complementary as 
measures of $V_{tb}$.

Both quark-antiquark annihilation and $W$-gluon fusion should be observed 
in Run II at the Tevatron, and yield a measurement of 
$V_{tb}$ with an accuracy of $\pm 0.1$.  Further running at the Tevatron
(30 fb$^{-1}$) could reduce the uncertainty to $\pm 0.05$.\cite{SmW,HBB}  
This is comparable to the accuracy on $V_{tb}$ 
expected from a measurement of the 
top-quark width at the $t\bar t$ threshold in $e^+e^-$ \cite{NLC} 
and $\mu^+\mu^-$~\cite{muon} colliders.  The accuracy with which $V_{tb}$
can be measured at the LHC via $W$-gluon fusion is limited only by 
systematics, most notably the uncertainty in the gluon distribution 
function.

\subsection{More on $W$-gluon fusion}

Let's look more closely at the calculation of the cross section for 
single-top-quark production via $W$-gluon fusion.  The leading-order 
cross section, shown in Fig.~4(a), is proportional to 
$\alpha_s\ln(m_t^2/m_b^2)$.  The large logarithm arises from the region 
of collinear $b\bar b$ production from the initial gluon.  Another power 
of this large logarithm appears at every order in perturbation theory 
via the emission of a collinear gluon from the internal $b$ quark, as 
shown in Fig.~4(b).  The resulting expansion parameter is thus 
$\alpha_s\ln(m_t^2/m_b^2)$, which yields a series which is much less 
convergent than an expansion in $\alpha_s$ alone.

\begin{figure}
\begin{center}
\epsfxsize= 4.25in
\leavevmode
\epsfbox{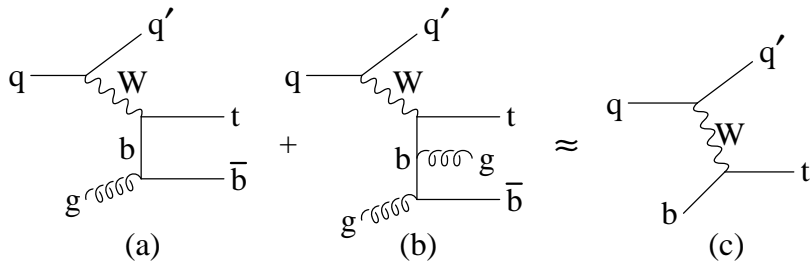}
\end{center}
\caption{Single-top-quark production via $W$-gluon at (a) leading order 
and (b) next-to-leading order.  The large logarithms, generated from the 
collinear region, can be summed into a $b$ distribution function via the 
DGLAP equations.  The leading-order diagram then becomes (c).}
\end{figure}

Fortunately, a technology exists to sum this large logarithm to all orders 
in perturbation theory.\cite{OT}  The idea is to use the DGLAP 
equations to sum the collinear logarithms.  In the process, one generates 
a perturbatively-derived $b$ distribution function, as shown in Fig.~4(c).  
The diagram in Fig.~4(c) becomes the leading-order diagram; it is 
of order $\alpha_s\ln(m_t^2/m_b^2)$, because the $b$ distribution 
function is intrinsically of this order.  The non-collinear part of the 
$W$-gluon fusion process (Fig.~4(a)) is only of order $\alpha_s$, so it 
corresponds to a correction to the leading-order process (Fig.~4(c)) of order
$1/\ln(m_t^2/m_b^2)$.  Virtual- and real-gluon emission corrections 
to Fig.~4(c) correspond to corrections of order $\alpha_s$.\cite{SSW} 

Both beauty and charm should have perturbatively-derivable distribution 
functions, since $m_b, m_c >> \Lambda_{\rm QCD}$.  This can be tested via 
heavy-flavor production at HERA.  For $Q^2 >> m_c^2$, the leading-order  
process for charm production is shown in Fig.~5, with a charm 
distribution function.\cite{ACOT}  The observed cross section should agree 
with that 
calculated via Fig.~5 (and radiative corrections to it), using a 
perturbatively-derived charm distribution function.  If it does not 
agree, then one must appeal to ``intrinsic charm'', which corresponds to 
a nonperturbative component to the charm distribution function.\cite{BHPS}
 
\begin{figure}
\begin{center}
\epsfxsize= 1.25in
\leavevmode
\epsfbox{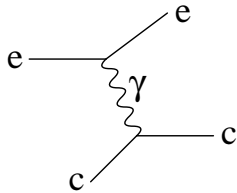}
\end{center}
\caption{Leading-order diagram for charm production at HERA for 
$Q^2 >> m_c^2$.}
\end{figure}

\section{$m_t$ and the $t\bar t$ threshold}

Although the top quark was only recently discovered, the top-quark mass 
is already the second best-known quark mass (as a percentage of its mass).
The top-quark mass is obtained from the invariant-mass distribution of its 
decay products, a $W$ and a $b$ jet.  The peak in this distribution 
corresponds closely with the top-quark pole mass.\footnote{I return to 
the top-quark pole mass at the end of this section.}  The CDF/D0 average is
\cite{N}
\begin{equation}
m_t=175.6 \pm 5.5 \; {\rm GeV} \;.
\end{equation}
This corresponds to an $\overline{\rm MS}$ mass of
\begin{equation}
\overline{m}_t(\overline{m}_t)=166.5 \pm 5.5 \; {\rm GeV} \;.
\end{equation}

In contrast, the bottom and charm
masses are obtained from the $\Upsilon$ and $\Psi$ spectra.  Since these 
spectra involve nonperturbative QCD, the most reliable method to obtain 
the quark mass is via lattice QCD.  This yields the $b$-quark mass 
\cite{NRQCD}
\begin{equation}
\overline{m}_b(m_b)=4.0 \pm 0.1 \; {\rm GeV} \;,
\end{equation}
which is the best-known quark mass.

Amongst the quarks, as well as the leptons, the top quark is unique;
it is the only fermion whose mass lies near the electroweak scale, which 
is the natural value for fermion masses in the standard model.  Fermions 
acquire their mass via their Yukawa coupling to the Higgs field, $y_f$:
\begin{equation}
m_f=y_f \frac{v}{\sqrt 2}	
\end{equation}
where $v\approx 250$ GeV is the vacuum-expectation value of the Higgs 
field.  The Yukawa coupling of the top quark is near unity, a natural value,
while all other fermions have a Yukawa coupling much less than unity.  
It therefore seems likely that, if we are ever to have a theory of fermion
masses, the top-quark mass will be the first to be understood.  This 
provides a motivation for measuring its mass as precisely as possible.

The electroweak gauge couplings have been measured to a precision of about
$0.1 \%$.  This provides a goal for the accuracy of the 
top-quark mass measurement, roughly $200$ MeV.\footnote{An example of a
theory of fermion masses in which a precision of 1 GeV on the top-quark mass
is sufficient is SO(10) grand unification with Yukawa-coupling 
unification.\cite{S}}   Such an accuracy cannot 
be achieved at hadron colliders; the best prospect is to make a precision 
measurement of the $t\bar t$ threshold in $e^+e^-$ or $\mu^+\mu^-$ 
colliders.  This is the analogue of the extraction of the bottom and 
charm masses from the $\Upsilon$ and $\Psi$ spectra.  However, due to the 
short top-quark lifetime, topononium states do not have time to form, so 
the $t\bar t$ threshold is rather different from that of bottom and charm.

\subsection{Toponium spectroscopy}

It is amusing to consider what toponium spectroscopy would be like if the 
top quark lived long enough to form toponium states.  The ground state 
would be very small, with a Bohr radius 
($C_F = 4/3$ is a color factor)
\begin{equation}
a = (C_F\frac{m_t}{2}\alpha_s)^{-1}\approx (12 \;{\rm GeV})^{-1}, 
\label{bohr}
\end{equation}
which is much less than the characteristic distance scale of 
nonperturbative QCD,
$\Lambda_{\rm QCD}^{-1} \approx (200 \;{\rm MeV})^{-1}$. 
The ground state, as well as the first few excited 
states, would be governed almost entirely by perturbative QCD, which 
gives rise to a potential
\begin{equation}
V(r)=-C_F\frac{\alpha_s(1/r)}{r}
\end{equation}
with a coupling that depends on the distance between the quark and the 
antiquark.  If we ignore the running of the coupling, we can use this 
Coulomb potential to estimate the number of toponium levels there would 
be below $T\overline T$ threshold ($T=t\bar q$ is a top meson).  This 
threshold is attained when the size of the toponium state is of order 
$\Lambda_{\rm QCD}^{-1}$.  Recall that the size of a Coulomb bound 
state is approximately $n^2a$, where $n$ is the principal quantum number.
Using Eq.~(\ref{bohr}) we obtain\footnote{The result that 
$n_{max} \sim m_t^{1/2}$ is actually true for an arbitrary 
potential.\cite{QR}}
\begin{equation}
n_{max} \approx 0.8 \left(\frac{m_t}{\Lambda_{\rm QCD}}\right)^{1/2}\;.
\label{nmax}
\end{equation}
The constant of proportionality in Eq.~(\ref{nmax}) was obtained by 
fitting this formula to the $\Upsilon$ spectrum, which has 
$n_{max}\approx 4$.\footnote{This includes the $\Upsilon(4s)$, which is 
only slightly above the $B\overline B$ threshold.  Eq.~(\ref{nmax}) also works 
well for the $\Psi$ spectrum, which has $n_{max} \approx 2$.}
We find that for toponium, the number of principal quantum numbers below 
threshold would be
\begin{equation}
n_{max}\approx 24\;.	
\end{equation}
Taking into account angular excitations, spin-orbit and spin-spin 
interactions, this corresponds to $2n_{max}^2\approx 1152$ distinct 
energy levels below threshold.

\subsection{$t\bar t$ threshold and the top-quark width}

In reality, toponium states do not have 
time to form before the top quark decays.  The formation time can be 
estimated by the size of the toponium state divided by the velocity of 
the quark and antiquark, which in a Coulomb bound state is approximately
$C_F\alpha_s$:
\begin{equation}
t_{form}\approx \frac{n^2a}{C_F\alpha_s}\approx n^2 (1.6 \;{\rm GeV})^{-1}\;.
\label{tform}
\end{equation}
[The formation time can also be estimated via 
$t_{form} \approx 1/|E_n|$, where 
\begin{equation}
E_n = - \frac{1}{2}\frac{m_t}{2}C_F^2\alpha_s^2/n^2
\label{en}
\end{equation} 
is the energy of the $n^{\rm th}$ level (relative to $2m_t$); 
this also yields Eq.~(\ref{tform}).]
The formation time is greater than the toponium lifetime, 
$(2\Gamma)^{-1}\approx 
(3 \;{\rm GeV})^{-1}$ (half the top-quark lifetime, since either the $t$ or
$\bar t$ can decay), for all but perhaps the ground state.  As a 
consequence, the $t\bar t$ threshold is a smooth continuum, with no sharp 
resonances, as shown in Fig.~6, curve (c).  The $1s$ resonance is pronounced if
we make the top quark width half its standard-model value, as shown in
curve (a) of Fig.~6.

\begin{figure}[ht]
\begin{center}
\vspace*{-1.25in}
\epsfxsize= 3in
\leavevmode
\epsfbox{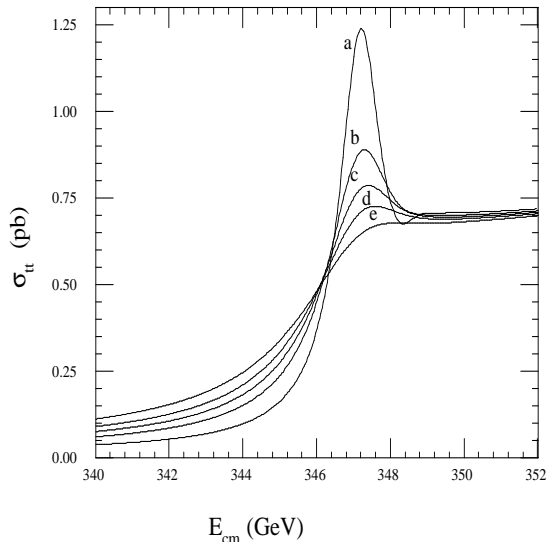}
\end{center}
\caption{The $t\bar t$ threshold in lepton collisions for $m_t=175$ GeV. 
The ratio of top quark width to the standard-model value is (a) 0.5, (b) 0.8,
(c) 1.0, (d) 1.2, (e) 1.5.  From Ref.~23.}
\end{figure}

The top-quark width plays a more important role than just smoothing out 
the $t\bar t$ threshold, however; it acts as an infrared cutoff, 
eliminating the contribution to the threshold from nonperturbative 
QCD.\cite{BDKKZ}  
One can see this in a heuristic way as follows.  Imagine that the 
top quark is stable, and consider the wave 
function of the $n^{\rm th}$ state, which falls off exponentially as
\begin{equation}
\psi_n \sim e^{-r/na}\sim e^{-r\sqrt{-m_tE_n}}
\label{psi}
\end{equation}
where I have used Eqs.~(\ref{bohr}) and (\ref{en}) to express the argument 
of the exponential in terms of the energy.  In reality 
there is not sufficient time for the bound states to form; one can implement
this information by saying that the bound states form in the complex energy 
plane at $E_n-i(2\Gamma)/2$.  To find the impact of these resonances on 
physics at real values of $E$, we let $E_n \to E_n + i\Gamma$ in 
Eq.~(\ref{psi}), to obtain
\begin{equation}
\psi_n \sim e^{-r\sqrt{-m_t(E_n+i\Gamma)}}
\label{psi2}
\end{equation}
In the absence of the width, the exponential suppression of the 
wave function at long distances is reduced as one approaches threshold,
$E_n \to 0$, and the toponium state has a large nonperturbative contribution 
from distances of order $\Lambda_{\rm QCD}^{-1}$ and greater.  However, with 
the width present, an exponential suppression remains even as one
approaches threshold:
\begin{equation}
|\psi_n| \stackrel{E_n\to 0}{\to} e^{-r\sqrt{m_t\Gamma/2}}\;.
\label{psi3}
\end{equation}
The width suppresses the wave function at distances greater than 
$1/\sqrt{m_t\Gamma}\approx
(16 \;{\rm GeV})^{-1}$, which is much less than $\Lambda_{\rm QCD}^{-1}
\approx (200\;{\rm MeV})^{-1}$.  Thus the 
$t\bar t$ threshold has a negligible contribution from 
nonperturbative QCD.  The threshold line shape, as shown in Fig.~6, can 
be calculated entirely within perturbation theory, 
limited in accuracy only by our strength to carry out higher-order 
calculations.

\subsection{Top-quark pole mass}

Because the $b$ quark is confined into hadrons, we are used to the notion that 
the $b$-quark pole mass is unphysical.  For example, one could attempt to
identify the $b$-quark pole mass as the mass of a $B$ meson minus the binding 
energy of the light quark, but this binding energy is an ambiguous concept,
due to confinement.  Since the binding energy
is of order $\Lambda_{\rm QCD}$, one expects an ambiguity in the $b$-quark pole
mass proportional to $\Lambda_{\rm QCD}$.

The above argument can be made more rigorous by studying the one-loop
contribution to the $b$-quark propagator, as shown in Fig.~7.  The region
of soft gluon momentum inevitably involves nonperturbative QCD.  It has been 
shown that this prevents an unambigous definition of the $b$-quark pole
mass, with an ambiguity proportional to $\Lambda_{\rm QCD}$.\cite{BSUV}  

\begin{figure}
\begin{center}
\epsfxsize= 1.5in
\leavevmode
\epsfbox{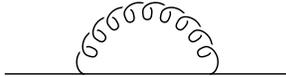}
\end{center}
\caption{One-loop contribution to the quark propagator.}
\end{figure}

Given that the width acts as an infrared cutoff in top physics, screening
the effects on nonperturbative QCD, one might expect the width to allow
an unambiguous definition of the top-quark pole mass.  This is not the case.
The width moves the perturbative top-quark pole into the complex energy plane,
but the position of the real part of the pole remains ambiguous by an amount
proportional to $\Lambda_{\rm QCD}$. The width does not remove the contribution
of soft gluons to the top-quark pole mass.\cite{pole}

For example, consider the measurement of the top-quark mass at hadron 
colliders by reconstructing the invariant mass of the decay products, a 
$W$ and a $b$ quark.  In reality, the $b$ quark manifests itself as a jet
of colorless hadrons.  Since the top quark is colored, at least one of the
quarks in the hadrons that make up the $b$ jet must not be a decay product
of the top quark.  Since there is no unambiguous way of subtracting the 
binding energy of this quark from the $b$ jet, 
the $Wb$ invariant mass is ambiguous by an amount of order 
$\Lambda_{\rm QCD}$.  Thus the top-quark
pole mass is ambiguous by an amount proportional to $\Lambda_{\rm QCD}$, 
despite the fact that the top quark does not form hadrons.

The moral is that if one hopes to measure the top-quark mass to an accuracy
of 200 MeV, one must use a short-distance mass, such as the $\overline
{\rm MS}$ mass, in the calculation.  There is no fundamental 
roadblock to measuring this mass to arbitrary precision from the 
$t\bar t$ threshold.

\section{Conclusions}

Because top-quark physics is free from the effects of nonperturbative QCD, 
we already 
have precision top-quark theory.  What is needed is precision top-quark
experiments.  We await a measurement of $V_{tb}$ to $\pm 0.1$ from 
single-top-quark production in Run II at the Tevatron.  Additional running
(30 fb$^{-1}$) could reduce the uncertainty to $\pm 0.05$.  The accuracy 
on $V_{tb}$ from $W$-gluon fusion at the LHC will be limited only by 
systematic errors, in particular the uncertainty in the gluon distribution 
function.  The top-quark $\overline{\rm MS}$ mass can be measured to 
200 MeV or even better
at $e^+e^-$ and $\mu^+\mu^-$ colliders running at the $t\bar t$ threshold.
By making precision studies of the top quark we hope not only to measure
the standard-model parameters accurately, but to uncover physics beyond the
standard model.

\section*{Acknowledgments}

I am grateful for conversations with A.~El-Khadra, T.~Liss, J.~Rosner, 
M.~Smith, T.~Stelzer, Z.~Sullivan, and W.-K.~Tung.
This work was supported in part by Department of Energy grant 
DE-FG02-91ER40677.

\end{document}